 \documentclass[pra, aps,  twocolumn, groupedaddress, showpacs, superscriptaddress]{revtex4-2}
 \usepackage{amssymb,amsmath,amsthm,color,graphicx,times,graphicx}
 \usepackage{hyperref}
 \usepackage{subfigure}
 \usepackage{times}
 \usepackage{bbold}
 \usepackage{color}

 \newcommand{\bra}[1]{\langle{#1}|}
 \newcommand{\ket}[1]{|{#1}\rangle}

 \providecommand{\openone}{\leavevmode\hbox{\small1\kern-4.3pt\normalsize1}}

 \theoremstyle{plain}

\theoremstyle{definition}

\usepackage{xcolor}

\begin{document}
 \title{Exploiting Non-Markovian Memory Effects for Robust Quantum Teleportation} 

\author{S. Gaidi}\email{safaegaidi19@gmail.com}  
\affiliation{LPHE-Modeling and Simulation, Faculty of Sciences, Mohammed V University in Rabat, Rabat, Morocco.}
\author{N.-E. Abouelkhir}\email{abouelkhir115@gmail.com}\affiliation{LPHE-Modeling and Simulation, Faculty of Sciences, Mohammed V University in Rabat, Rabat, Morocco.}
 \author{A. Slaoui}\email{abdallah.slaoui@um5s.net.ma}\affiliation{LPHE-Modeling and Simulation, Faculty of Sciences, Mohammed V University in Rabat, Rabat, Morocco.}\affiliation{Centre of Physics and Mathematics, CPM, Faculty of Sciences, Mohammed V University in Rabat, Rabat, Morocco.}
 \author{M. El Falaki}\affiliation{LPHE-Modeling and Simulation, Faculty of Sciences, Mohammed V University in Rabat, Rabat, Morocco.}\affiliation{ Laboratory of Innovation in Science, Technology and Modeling, Faculty of Sciences of El Jadida, Chouaib Doukali University, El Jadida, Morocco}
 \author{R. Ahl Laamara}\affiliation{LPHE-Modeling and Simulation, Faculty of Sciences, Mohammed V University in Rabat, Rabat, Morocco.}\affiliation{Centre of Physics and Mathematics, CPM, Faculty of Sciences, Mohammed V University in Rabat, Rabat, Morocco.}

\begin{abstract}

The reliable transmission of quantum information remains a central challenge in the presence of environmental noise. In particular, maintaining high teleportation fidelity in open quantum systems is hindered by decoherence, which disrupts quantum coherence and entanglement. Traditional noise mitigation techniques often neglect the rich temporal correlations present in realistic environments. This raises a key question: can non-Markovian memory effects be harnessed to improve the performance of quantum teleportation? In this work, we address this problem by analyzing how non-Markovian dynamics influence teleportation fidelity. We employ a statistical speed approach based on the Hilbert Schmidt norm to witness information backflow and monitor the system's instantaneous evolution rate. Our study focuses on two measurement-based strategies: weak measurement (WM) combined with quantum measurement reversal (QMR), and a hybrid protocol integrating environment-assisted measurement (EAM) with post-selection and QMR.
Through analytical expressions and detailed numerical simulations, we demonstrate that both strategies can enhance teleportation fidelity under non-Markovian noise. Notably, the EAM-based scheme exhibits superior robustness, achieving high fidelity even without fine-tuned parameters. Our results establish a concrete link between non-Markovian memory effects, statistical speed, and coherence preservation, offering practical insights for the design of resilient quantum communication protocols.

 \end{abstract}
 	\date{\today}

\maketitle
\section{Introduction}

Quantum teleportation is a fundamental protocol in quantum information science that allows the transmission of an arbitrary quantum state between two distant parties by leveraging quantum entanglement and classical communication \cite{Bennett1993, Bouwmeester1997, Pirandola2015, Kimble2008, Wehner2018, Gao2010, Ursin2007}. This process is essential to the realization of quantum communication networks and distributed quantum computation \cite{Briegel1998, Cirac1997, VanMeter2014, Northup2014, Gisin2007, Slaoui2023a, Slaoui2024}. However, in practical implementations, ideal isolation from environmental noise is unachievable. Decoherence \cite{Zurek2003, Breuer2002, Schlosshauer2007}, induced by system-environment interactions, significantly degrades the fidelity of quantum information protocols. These interactions can often destroy entanglement \cite{Yu2004, Almeida2007, Bellomo2007, Kirdi2023a, Kirdi2023b}, reducing the effectiveness of teleportation and necessitating the development of robust strategies for preserving coherence \cite{Lidar2014, Nielsen2010, Preskill2018}.\par

In realistic settings, quantum systems are modeled as open systems. Their dynamics are typically non-unitary, influenced by their interaction with an external environment. Depending on the nature of this interaction, the evolution can be either  Markovian, where information lost to the environment is not recovered, or non-Markovian, where a temporary backflow of information occurs \cite{Breuer2009, Rivas2010, Jahromi2020, Deffner2013, Addis2014}. The latter allows for the partial restoration of lost coherence and correlations, making non-Markovianity a valuable feature in quantum control and communication \cite{Korotkov2006, Kim2012, Zhou2020, Liu2011, Chin2012}.\par

Quantifying non-Markovianity has become an area of intense research \cite{Chruściński2014, Vacchini2011, Gaidi2025, Abouelkhir2024}. Several approaches have been proposed, including measures based on distinguishability of quantum states \cite{Laine2010}, divisibility of dynamical maps \cite{Chruściński2011}, and geometric characteristics of quantum trajectories \cite{Wang2009}. One particularly promising approach is the use of the Hilbert–Schmidt speed (HSS), which captures the instantaneous rate of change of a quantum state in Hilbert space. The sign of the time derivative of HSS can serve as a witness for non-Markovianity: a positive derivative signals a backflow of information, indicating a non-Markovian interval. Compared to other geometric measures, HSS is computationally simpler and more scalable to high dimensional systems \cite{Jahromi2020, Abouelkhir2024, Modi2012, Guarnieri2016}.\par

Recent experimental advances have demonstrated that quantum error correction can significantly extend the coherence time of quantum states, as evidenced in superconducting qubits \cite{Ofek2016, Andersen2020, Kelly2015}. However, such error correction is technically demanding. An alternative and complementary class of techniques involves measurement-based control. Weak measurement (WM), enables the partial collapse of a quantum state without fully destroying its coherence \cite{Abouelkhir2023, Abdellaoui2024, Abouelkhir2025}. When combined with a quantum measurement reversal (QMR), the overall protocol allows for probabilistic recovery of the premeasurement state, even after environmental decoherence \cite{Korotkov2006, Kim2012, Li2013}.\par

Some other works  showed that quantum feedback based on weak measurement could significantly mitigate decoherence in solid-state qubits \cite{Korotkov2006, Kim2012, Ashhab2010}. Theoretical studies \cite{Jahromi2020, Wang2009} and experiments \cite{Kim2012, Katz2008} have confirmed that the WM+QMR strategy enhances the survival of quantum coherence in noisy environments. More recently, environment-assisted measurement (EAM) schemes have emerged, where measurement on the environment selectively filters out strongly decohered trajectories. Combined with QMR, this can further enhance fidelity \cite{Zhou2020, Abouelkhir2024}.\par

In this paper, we investigate the role of non-Markovian dynamics in enhancing quantum teleportation fidelity and assess the performance of two measurement-based strategies: WM+QMR and EAM+QMR. We aim to establish how these techniques interact with the memory effects inherent to non-Markovian environments and whether this interaction can be harnessed to preserve teleportation fidelity.\par

To quantify the dynamics, we use teleportation fidelity as a benchmark and compute the Hilbert–Schmidt speed to witness non-Markovianity. Our analysis shows that memory effects, as indicated by positive values of the HSS derivative, correspond closely to revivals in teleportation fidelity. This correlation confirms that non-Markovianity can be beneficial for communication tasks if correctly exploited.\par

Furthermore, we demonstrate that while non-Markovianity can provide temporary protection against decoherence, optimal preservation of quantum fidelity is achieved when it is combined with WM+QMR or EAM+QMR. The WM+QMR scheme performs better when weak measurement strength is finely tuned but is sensitive to parameter variations. The EAM+QMR scheme, by contrast, is more robust and consistently improves fidelity.\par

These findings reinforce the idea that quantum information protection benefits from a hybrid strategy leveraging both environmental memory and active control. Our work contributes to ongoing efforts to develop scalable, realistic solutions for preserving quantum information in open systems, offering potential applications in future quantum networks.\par

This paper is organized as follows. In Section II, we introduce the concept of quantum evolution speed using the Hilbert–Schmidt metric and discuss its relevance in open quantum systems. Section III is devoted to a non-Markovianity measure based on the Hilbert–Schmidt speed, which is used to detect memory effects in quantum channels. In Section IV, we analyze the behavior of teleportation fidelity under non-Markovian environments and establish the correlation with Hilbert–Schmidt speed. Section V proposes a control protocol combining WM and QMR to enhance teleportation fidelity. Section VI introduces an alternative strategy based on EAM and QMR, and compares its performance to the WM-based approach. Finally, Section VII summarizes the main results and outlines potential directions for future research.

\section{Quantum Evolution Speed Based on the Hilbert-Schmidt Metric}
\label{setion2}
A fundamental question in quantum information and open quantum system theory is how to quantify the speed at which a quantum state changes under a given physical evolution. This question becomes particularly relevant when studying decoherence, dissipation, and the influence of non-Markovian environments. Among various frameworks used to characterize the dynamics of quantum systems, the notion of quantum statistical speed has emerged as a versatile and insightful tool. This quantity captures how fast a quantum state evolves along a trajectory defined by a family of parameter-dependent states. In this context, the HSS constitutes a computationally favorable variant that is especially suited for witnessing memory effects without requiring spectral decompositions or eigenvalue computations.
In classical statistics, one often measures the dissimilarity between two probability distributions \( p = \{p_x\} \) and \( q = \{q_x\} \) through distance measures of the form \cite{Jahromi2020, abouelkhir2023estimating}:
\begin{equation}
    d_\alpha(p, q) = \frac{1}{2} \sum_x |p_x - q_x|^\alpha, \qquad \alpha \geq 1,
\end{equation}
where \( x \) denotes the possible outcomes of a random variable. This family of metrics obeys fundamental distance axioms such as positivity, symmetry, and the triangle inequality. When extended to a continuous variable, the summation is naturally replaced by an integral.

To generalize this concept to the quantum regime, one considers quantum states \( \rho \) and \( \sigma \) and associates them with outcome probabilities via a positive operator-valued measure (POVM) \( \{E_x\} \) satisfying \( E_x \geq 0 \) and \( \sum_x E_x = \mathbb{I} \). The outcome probabilities are given by \( p_x = \mathrm{Tr}[E_x \rho] \) and \( q_x = \mathrm{Tr}[E_x \sigma] \). By optimizing over all POVMs, the quantum analog of \( d_\alpha \) is defined as:
\begin{equation}
    D_\alpha(\rho, \sigma) = \max_{\{E_x\}} d_\alpha(p, q).
\end{equation}
For \( \alpha = 1 \), this corresponds to the wellknown trace distance :
\begin{equation}
    D_{\mathrm{Tr}}(\rho, \sigma) = \frac{1}{2} \mathrm{Tr}|\rho - \sigma|,
\end{equation}
which is contractive under completely positive trace-preserving (CPTP) maps and widely used in quantifying distinguishability and detecting non-Markovianity.

When \( \alpha = 2 \), one obtains the Hilbert-Schmidt distance:
\begin{equation}
    D_{\mathrm{HS}}(\rho, \sigma) = \sqrt{\frac{1}{2} \mathrm{Tr}[(\rho - \sigma)^2]}.
\end{equation}
This metric arises naturally from the Hilbert-Schmidt norm and benefits from an algebraic structure that facilitates analytical calculations. However, unlike the trace distance, the Hilbert-Schmidt metric is not contractive in general, particularly for systems with Hilbert space dimension greater than two. This poses limitations for its direct use as a non-Markovianity witness. Nevertheless, it remains a powerful auxiliary quantity, especially in low-dimensional systems like qubits and qutrits.

To assess how fast a quantum state \( \rho(\varphi) \) changes under variation of a real parameter \( \varphi \) (e.g., time, phase, coupling strength), one can define a quantum statistical speed based on a chosen distance function. For general \( \alpha \), the quantum statistical speed is given by \cite{abouelkhir2023estimating}:
\begin{equation}
    S_\alpha[\rho(\varphi)] = \left( \frac{1}{2} \mathrm{Tr} \left| \frac{d\rho(\varphi)}{d\varphi} \right|^\alpha \right)^{1/\alpha}.
\end{equation}
In this expression, the derivative \( \frac{d\rho(\varphi)}{d\varphi} \) captures the instantaneous rate of change of the state, and the norm characterizes the “length” of this change in state space.\par

Particularly, for \( \alpha = 2 \), one obtains the HSS, which takes the simple form:
\begin{equation}
    \mathrm{HSS}[\rho(\varphi)] = \sqrt{\frac{1}{2} \mathrm{Tr}\left[ \left( \frac{d\rho(\varphi)}{d\varphi} \right)^2 \right]}.
    \label{eq:HSS_def}
\end{equation}
This expression quantifies the instantaneous rate at which the quantum state changes in the Hilbert-Schmidt sense. Notably, the HSS does not involve the spectral decomposition of the state or its derivative, making it computationally efficient even for time dependent dynamics. This contrasts with other metrics like the Bures distance or quantum Fisher information, which often require diagonalization.

\section{HSS-Based Non-Markovianity Measure}

It is well established that non-Markovian memory effects can accelerate quantum evolution, revealing a connection between the speed of quantum dynamics and information backflow \cite{Liu2016, Mirkin2016, Ahansaz2019, Zou2020, Gaidi2025, Gaidi2024}. Among various approaches proposed to capture such effects, one promising method involves using quantum statistical speeds as witnesses of non-Markovianity. In particular, the HSS, a contractive and computationally convenient quantity, has been identified as a useful indicator of non-Markovian behavior, especially in low dimensional quantum systems \cite{Jahromi2020}.

Following the framework introduced in \cite{Jahromi2020}, we consider the HSS-based witness \( \chi(t) \), which quantifies the instantaneous speed of evolution of a quantum state. Positive values of \( \chi(t) \) indicate an increase in the evolution speed, which is typically associated with an information backflow a hallmark of non-Markovian dynamics.

To formally define this measure, we consider a quantum system with an $n$-dimensional Hilbert space \( \mathcal{H} \). The initial state is taken as a uniform superposition:
\begin{equation}
|\psi_0\rangle = \frac{1}{\sqrt{n}}( e^{i \varphi}  |\psi_1\rangle+|\psi_2\rangle+....+|\psi_n\rangle ),
\end{equation}
where, \( \varphi \) is an unknown phase shift and the set $\{|\psi_1\rangle,...,|\psi_n \rangle\}$ constitutes a complete orthonormal basis for the Hilbert space \( \mathcal{H} \).\par

The instantaneous HSS-based witness is then defined as
\begin{equation}
\chi(t) := \partial_t \text{HSS}(\rho_\varphi(t)),
\end{equation}
and the degree of non-Markovianity over a given time interval is quantified as:
\begin{equation}
\mathcal{N}_{\text{HSS}} := \max_{\varphi, \{|\psi_1\rangle, \dots, |\psi_n\rangle\}} \int_{\chi(t) > 0} \chi(t)\, dt,
\end{equation}
where the maximization is formally taken over all parameterizations of the initial state and evolution parameters, although in this work we do not perform such optimization. Our goal is to assess whether the HSS-based witness is capable of detecting non-Markovian behavior, rather than determining its maximum value.

This method offers practical advantages, notably avoiding the need for diagonalization of the system's density matrix. This feature becomes particularly valuable in the study of high-dimensional or multipartite open quantum systems. Moreover, among the family of quantum statistical speeds parameterized by \( \alpha \) (as discussed in Ref.\cite{abouelkhir2023estimating}, the choice \( \alpha = 2 \), corresponding to the Hilbert-Schmidt speed, provides a compelling balance between analytical tractability and sensitivity to non-Markovian features.

\section{Teleportation of Quantum State Fidelity under Non-Markovian Environment}

\begin{figure}
    \centering
    \includegraphics[width=1\linewidth]{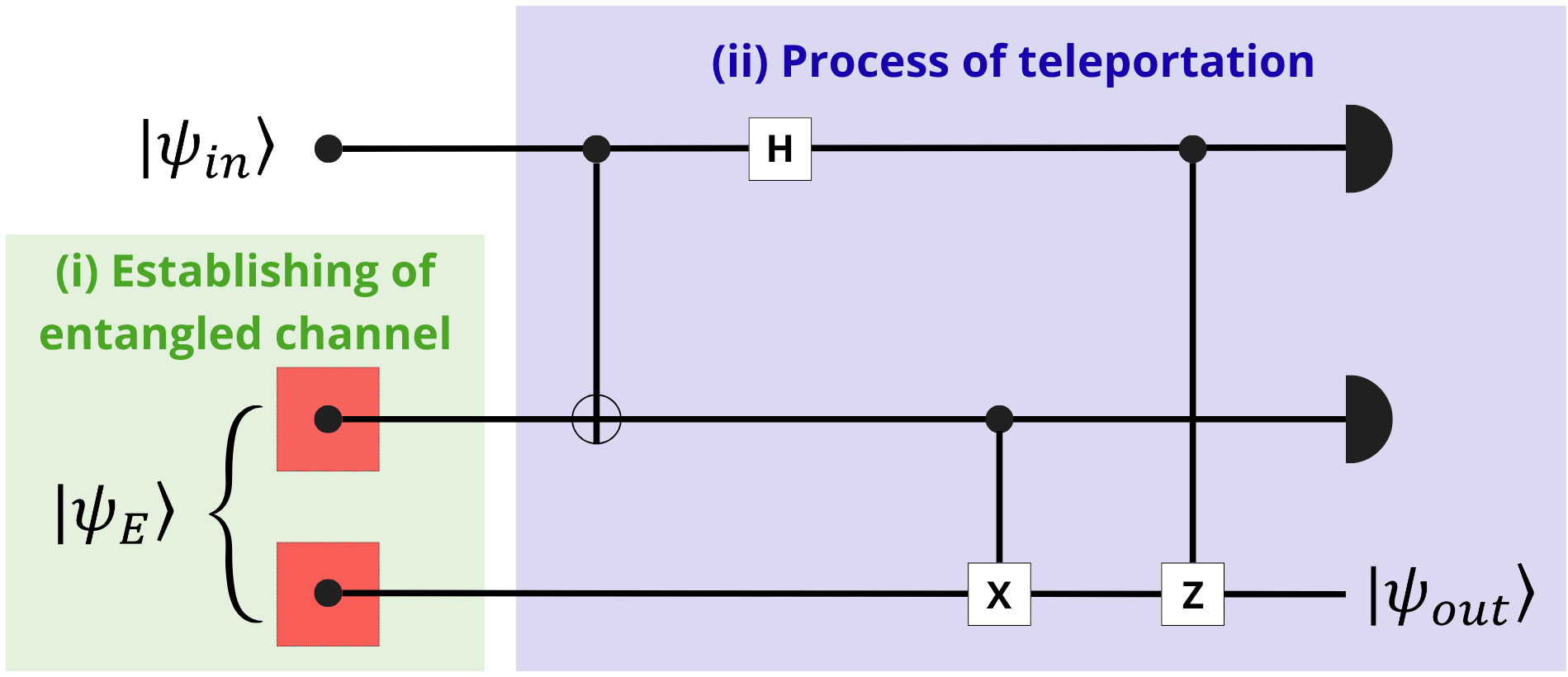}
    \caption{A quantum circuit for teleportation under non-Markovian noise, illustrating the protocol in two parts: (i) the establishment of a maximally entangled state between Qubits 2 and 3 subject to a non-Markovian noisy environment and (ii) the teleportation process. The top two qubits belong to Alice, and the bottom one to Bob. Quantum measurements are indicated by ammeter symbols, and the dotted boxes highlight the regions affected by noise.}
    \label{fig:enter-label}
\end{figure}

We consider the teleportation of a single-qubit state containing an unknown phase parameter $\varphi$, given by:
\begin{equation}
|\psi\rangle_1 = \cos\frac{\theta}{2}|0\rangle_1 + \sin\frac{\theta}{2}e^{i\varphi}|1\rangle_1.
\end{equation}

The quantum channel used for teleportation is initially prepared as a maximally entangled Bell state:
\begin{equation}
|\psi^+\rangle_{23} = \frac{1}{\sqrt{2}}(|00\rangle_{23} + |11\rangle_{23}),
\end{equation}
where qubits 2 and 3 are sent to Alice and Bob, respectively, through independent non-Markovian noisy environments.

Each qubit interacts resonantly with its local environment characterized by a Lorentzian spectral density:
\begin{equation}
J(\omega) = \frac{1}{2\pi}\frac{\gamma_0 \lambda^2}{(\omega - \omega_0)^2 + \lambda^2},
\end{equation}
where $\omega_0$ is the transition frequency of the qubit, $\gamma_0$ is the decay rate in the Markovian limit, and $\lambda$ describes the spectral width related to the memory effects of the environment.

Assuming the environments are initially in the vacuum state, the qubit dynamics can be described by the following Kraus operators \cite{Haseli2014, Li2023}:
\begin{equation}
K_0(t) = \begin{pmatrix}
1 & 0 \\
0 & \sqrt{\mu(t)}
\end{pmatrix}, \quad
K_1(t) = \begin{pmatrix}
0 & \sqrt{1 - \mu(t)} \\
0 & 0
\end{pmatrix},
\end{equation}
where the parameter $\mu(t)$ quantifies the population of the excited state and is given by:
\begin{equation}
\mu(t) = e^{-\lambda t}\left[\cos\left(\frac{d t}{2}\right) + \frac{\lambda}{d}\sin\left(\frac{d t}{2}\right)\right]^2,
\end{equation}
with $d = \sqrt{2\gamma_0\lambda - \lambda^2}$.

The evolution of the entangled state under the local non-Markovian channels is given by:
\begin{equation}
\rho(t) = \sum_{ij} K_i(t) \otimes K_j(t) \, \rho_0 \, K_i^\dagger(t) \otimes K_j^\dagger(t),
\end{equation}
where $\rho_0 = |\psi^+\rangle_{23}\langle\psi^+|$.

Assuming both qubits undergo identical non-Markovian noise, the resulting shared state between Alice and Bob becomes:
\begin{equation}
\begin{aligned}
\rho(t) &= f_{11}|00\rangle\langle 00| + f_{14}|00\rangle\langle 11| + f_{22}|01\rangle\langle 01| \\
&\quad + f_{33}|10\rangle\langle 10| + f_{41}|11\rangle\langle 00| + f_{44}|11\rangle\langle 11|,
\end{aligned}
\end{equation}
where the coefficients are given by:
\begin{equation}
\begin{aligned}
f_{11} &= \frac{\mu^2(t) - 2\mu(t) + 2}{2}, \\
f_{14} &= f_{41}^* = \frac{\mu(t)}{2}, \\
f_{22} &= f_{33} = \frac{\mu(t) - \mu^2(t)}{2}, \\
f_{44} &= \frac{\mu^2(t)}{2}.
\end{aligned}
\end{equation}


Through the teleportation protocol, the final state received by Bob is given by:
\begin{equation}
\begin{aligned}
\rho_{\text{out}}(t) &= \varepsilon_{11}(t)|0\rangle\langle 0| + \varepsilon_{22}(t)|1\rangle\langle 1| \\
&\quad + \varepsilon_{12}(t)|0\rangle\langle 1| + \varepsilon_{21}(t)|1\rangle\langle 0|,
\end{aligned}
\end{equation}
where the elements are given by:
\begin{equation}
\begin{aligned}
\varepsilon_{11}(t) &= \frac{1 + \mu(t)}{2} \cos^2\frac{\theta}{2} + \frac{1 - \mu(t)}{2} \sin^2\frac{\theta}{2}, \\
\varepsilon_{22}(t) &= \frac{1 - \mu(t)}{2} \cos^2\frac{\theta}{2} + \frac{1 + \mu(t)}{2} \sin^2\frac{\theta}{2}, \\
\varepsilon_{12}(t) &= \frac{\mu(t)}{2} \sin\theta\, e^{-i\varphi}, \\
\varepsilon_{21}(t) &= \frac{\mu(t)}{2} \sin\theta\, e^{i\varphi}.
\end{aligned}
\end{equation}

Thus, the output state can be explicitly rewritten as:
\begin{equation}
\begin{aligned}
\rho_{\text{out}}(t) &= \left[\frac{1 + \mu(t)}{2} \cos^2\frac{\theta}{2} + \frac{1 - \mu(t)}{2} \sin^2\frac{\theta}{2}\right] |0\rangle\langle 0| \\
&\quad + \left[\frac{1 - \mu(t)}{2} \cos^2\frac{\theta}{2} + \frac{1 + \mu(t)}{2} \sin^2\frac{\theta}{2}\right] |1\rangle\langle 1| \\
&\quad + \frac{\mu(t)}{2} \sin\theta\, e^{-i\varphi}|0\rangle\langle 1| + \frac{\mu(t)}{2} \sin\theta\, e^{i\varphi}|1\rangle\langle 0|.
\end{aligned}
\end{equation}

Using the above expressions, the fidelity between the input state $|\psi\rangle$ and the output state $\rho_{\text{out}}(t)$ is calculated as:
\begin{equation}
\mathcal{F}(t) = \langle\psi|\rho_{\text{out}}(t)|\psi\rangle.
\end{equation}

\begin{widetext}

 		\begin{figure}[hbtp]
 			{{\begin{minipage}[b]{.33\linewidth}
 						\centering
 						\includegraphics[scale=0.33]{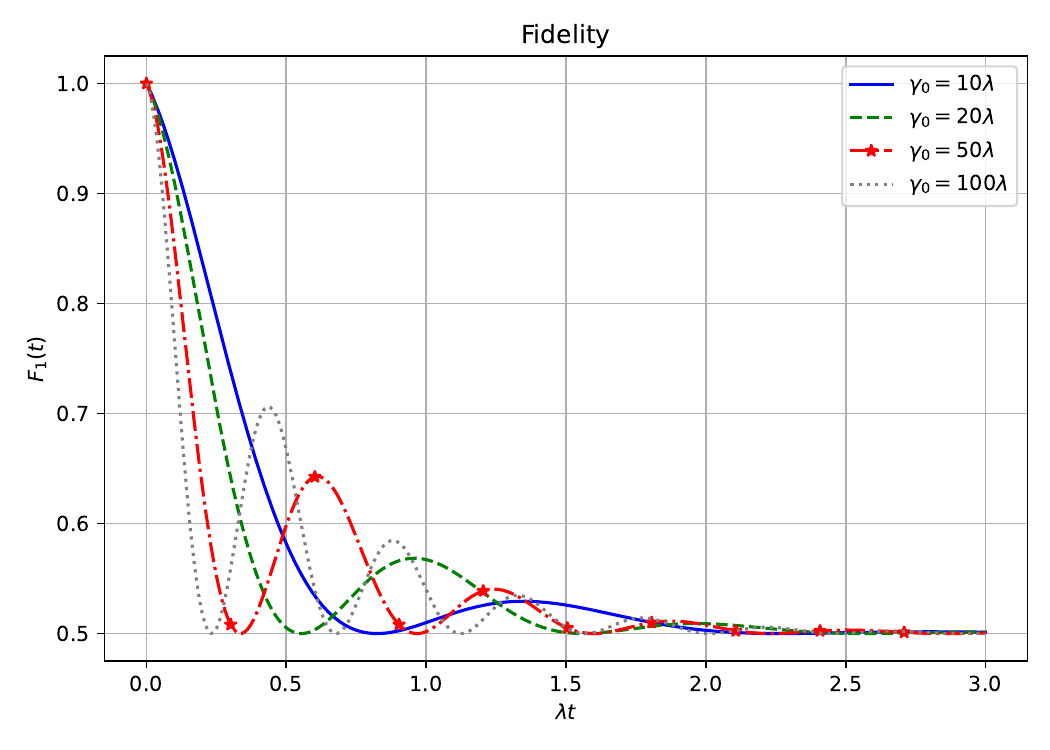} \vfill $\left(a\right)$
 					\end{minipage} \hfill
 					\begin{minipage}[b]{.33\linewidth}
 						\centering
 						\includegraphics[scale=0.33]{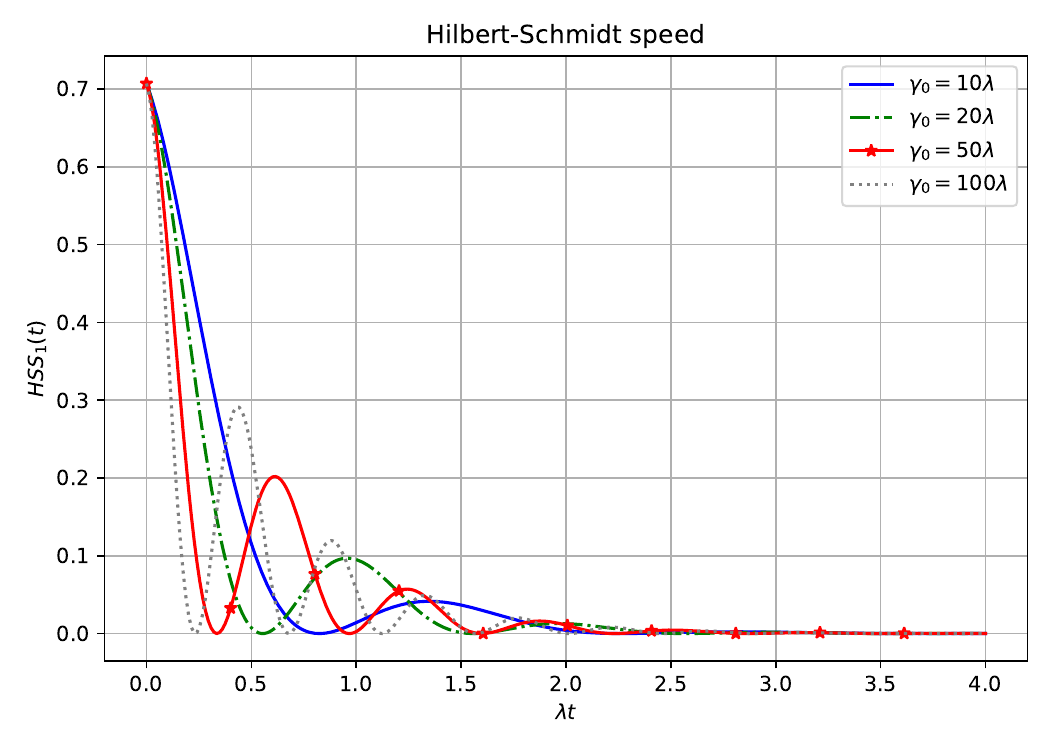} \vfill  $\left(b\right)$
 					\end{minipage} \hfill
 					\begin{minipage}[b]{.33\linewidth}
 						\centering
 						\includegraphics[scale=0.33]{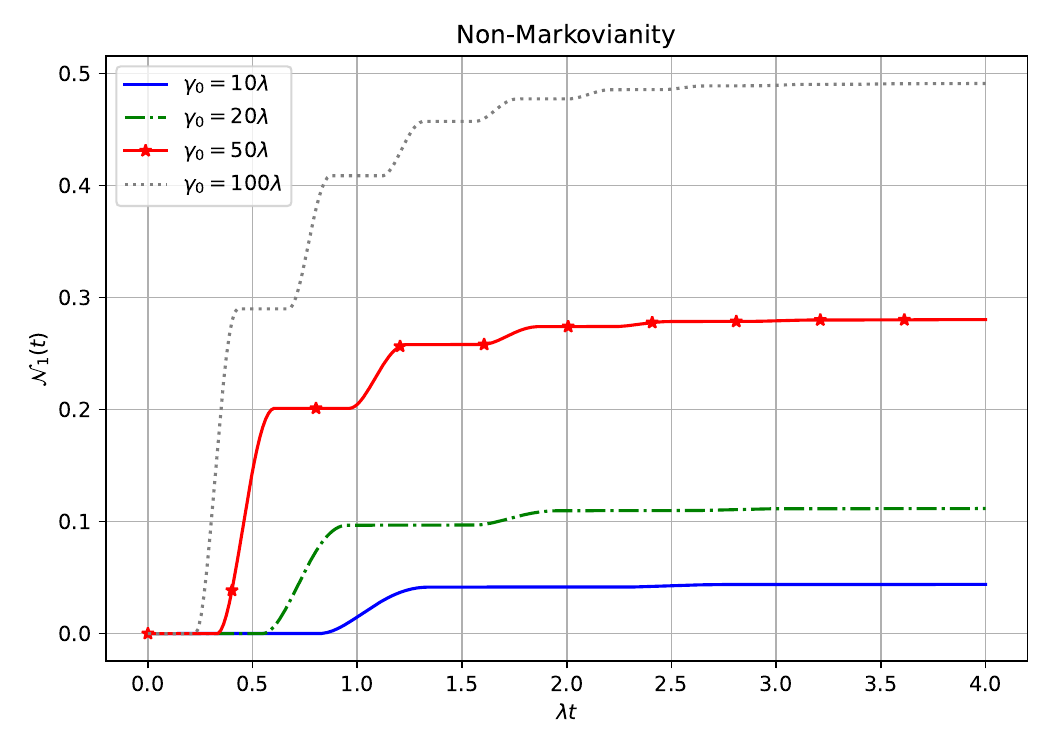} \vfill  $\left(c\right)$
 			\end{minipage}}}
             \caption{Teleportation fidelity $\mathcal{F}_1(t)$ , $HSS_1(t)$ and Non Markovianity $\mathcal{N}_1(t)$  versus $\lambda$t  with different coupling strengths $\gamma_0 = 10\lambda$, $\gamma_0 = 20\lambda$,$\gamma_0 = 50\lambda$ and $\gamma_0 = 100\lambda$. The initial state parameter is set to $\theta = \frac{\pi}{2}$.}
\label{fig1}
 		\end{figure}
 	\end{widetext}

After substituting the expressions of the matrix elements and simplifying, the fidelity becomes:
\begin{equation}
\begin{aligned}
\mathcal{F}_1(t) &= \langle\psi|\rho_{\text{out}}(t)|\psi\rangle \\
&= \varepsilon_{11}(t) \cos^2\frac{\theta}{2} + \varepsilon_{22}(t) \sin^2\frac{\theta}{2} \\
&\quad + \sin{\theta}  Re[\varepsilon_{12}(t)  e^{i\varphi}].
\end{aligned}
\end{equation}
Figure~\ref{fig1} illustrates the behavior of the teleportation fidelity $\mathcal{F}_1(t)$, the Hilbert–Schmidt speed $HSS_1(t)$, and the cumulative non-Markovianity $\mathcal{N}_1(t)$ as functions of $\lambda t$, with the input state initialized at $\theta = \frac{\pi}{2}$, representing a maximally coherent superposition. In panel (a), the fidelity exhibits damped oscillations followed by a gradual decay, signaling the influence of non-Markovian memory effects. These oscillations indicate temporary information backflow from the environment to the system, which momentarily restores quantum coherence and enhances teleportation performance. However, this revival is not sufficient to completely counteract the long term impact of decoherence, as the fidelity ultimately decreases due to accumulated noise. The degradation becomes more pronounced under stronger coupling (i.e., smaller $\gamma_0/\lambda$), highlighting the limitations of passive memory effects. Panel (b) displays the Hilbert–Schmidt speed, which captures the instantaneous rate of change in the system's quantum state, peaks in $HSS_1(t)$ coincide with fidelity revivals and mark intervals of enhanced dynamical activity driven by environmental memory. This reinforces the presence of non-Markovian dynamics and indicates moments when the system’s evolution is accelerated by coherence recovery. Panel (c) shows the cumulative non-Markovianity $\mathcal{N}_1(t)$, which increases in discrete steps, each corresponding to an information backflow. The stepwise growth reaching values near 0.5 confirms that while the environment provides temporary protection via memory effects, it cannot indefinitely preserve quantum coherence. Overall, the correlation between $\mathcal{F}_1(t)$, $HSS_1(t)$, and $\mathcal{N}_1(t)$ demonstrates that although non-Markovianity contributes to fidelity revival and faster state evolution, it alone is insufficient to sustain high teleportation fidelity in the long time limit, thereby motivating the implementation of active control techniques such as weak measurement reversal and environment-assisted measurement to extend coherence protection.
\section{Optimizing Teleportation Fidelity via Weak Measurement and Measurement Reversal}

To mitigate the degradation of teleportation fidelity caused by non-Markovian noise, we propose a robust three-step protocol that integrates WM and QMR. This strategy is designed to protect the entanglement resource throughout the teleportation process and enhance the fidelity of the teleported state.

Immediately after preparing the maximally entangled state, Charlie performs weak measurements on qubits 2 and 3. The action of a weak measurement with strength $p_j$ on the $j$th qubit is described by \cite{Li2023}:
\begin{equation}
|0\rangle_j \rightarrow |0\rangle_j, \quad |1\rangle_j \rightarrow \sqrt{1 - p_j}\, |1\rangle_j,
\end{equation}
where $0 \leq p_j \leq 1$ and $j = 2,3$. Assuming symmetric measurements ($p_2 = p_3 = p$), the shared entangled state becomes:
\begin{equation}
|\Psi^+\rangle^{\text{WM}}_{23} = \frac{1}{\sqrt{2W}}\left(|00\rangle_{23} + \bar{p}\, |11\rangle_{23}\right),
\end{equation}
where $\bar{p} = 1 - p$ and $W = \frac{1}{2}(1 + \bar{p}^2)$. This operation effectively skews the state toward the ground state $|00\rangle$, thereby making it more robust against decoherence.

The weakly measured state is then sent through two independent local non-Markovian channels to Alice and Bob. Under the influence of environmental noise described by a function $\mu(t)$, the shared state evolves into:
\begin{align}
\rho_1(t) = \frac{1}{W} \Big(& 
    \zeta_{11} \ket{00}\bra{00} + \zeta_{14} \ket{00}\bra{11} + \zeta_{22} \ket{01}\bra{01} \notag \\
    & + \zeta_{33} \ket{10}\bra{10} + \zeta_{41} \ket{11}\bra{00} + \zeta_{44} \ket{11}\bra{11} 
\Big),
\end{align}
where the time-dependent coefficients are:
\begin{align}
\zeta_{11} &= \frac{1}{2} + \frac{1}{2} \bar{p}^2(1 - \mu(t))^2, \\
\zeta_{14} &= \zeta_{41}^* = \frac{1}{2} \bar{p} \mu(t), \\
\zeta_{22} &= \zeta_{33} = \frac{1}{2} \bar{p}^2 (\mu(t) - \mu^2(t)), \\
\zeta_{44} &= \frac{1}{2} \bar{p}^2 \mu^2(t).
\end{align}

After receiving their qubits, Alice and Bob apply local quantum measurement reversal operations. A QMR with strength $q_j$ on the $j$th qubit is defined as:
\begin{equation}
|0\rangle_j \rightarrow \sqrt{1 - q_j}\, |0\rangle_j, \quad |1\rangle_j \rightarrow |1\rangle_j,
\end{equation}
where $0 \leq q_j \leq 1$ and we assume $q_2 = q_3 = q$. After applying QMR, the state becomes:
\begin{align}
\rho_2(t) &= \frac{1}{V} \Big(
    q^2 \zeta_{11} |00\rangle\langle 00| 
    + \bar{q} \zeta_{14} |00\rangle\langle 11| 
    + \bar{q} \zeta_{22} |01\rangle\langle 01| \notag \\
    &\quad + \bar{q} \zeta_{33} |10\rangle\langle 10| 
    + \bar{q} \zeta_{41} |11\rangle\langle 00| 
    + \zeta_{44} |11\rangle\langle 11|
\Big),
\end{align}

with $\bar{q} = 1 - q$ and normalization constant:
\begin{equation}
V = \bar{q}^2 \zeta_{11} + \bar{q} \zeta_{22} + \bar{q} \zeta_{33} + \zeta_{44}.
\end{equation}

Following the standard teleportation protocol, the state received by Bob becomes:
\begin{equation}
\rho'_{\text{out}} = \begin{pmatrix}
\chi_{11}(t) & \chi_{12}(t) \\
\chi_{21}(t) & \chi_{22}(t)
\end{pmatrix},
\end{equation}
where the elements are given by:
\begin{align}
\chi_{11}(t) &= \frac{1}{V}\left( \bar{q}^2 \zeta_{11} + \zeta_{44} \right) \cos^2\frac{\theta}{2} + \frac{2}{V} \bar{q} \zeta_{22} \sin^2\frac{\theta}{2}, \\
\chi_{12}(t) &= \frac{1}{V} \bar{q} \zeta_{14} \sin\theta\, e^{-i\varphi}, \\
\chi_{21}(t) &= \frac{1}{V} \bar{q} \zeta_{14} \sin\theta\, e^{i\varphi}, \\
\chi_{22}(t) &= \frac{2}{V} \bar{q} \zeta_{22} \cos^2\frac{\theta}{2} + \frac{1}{V} \left( q^2 \zeta_{11} + \zeta_{44} \right) \sin^2\frac{\theta}{2}.
\end{align}
The teleportation fidelity is then:
\begin{equation}
\mathcal{F}_2(t) = \langle\psi|\rho'_{\text{out}}(t)|\psi\rangle = \chi_{11}(t) \cos^2\frac{\theta}{2} + \sin\theta\, \mathrm{Re}[\chi_{12}(t) e^{i\varphi}].
\end{equation}
\begin{widetext}

 		\begin{figure}[hbtp]
 			{{\begin{minipage}[b]{.33\linewidth}
 						\centering
 						\includegraphics[scale=0.33]{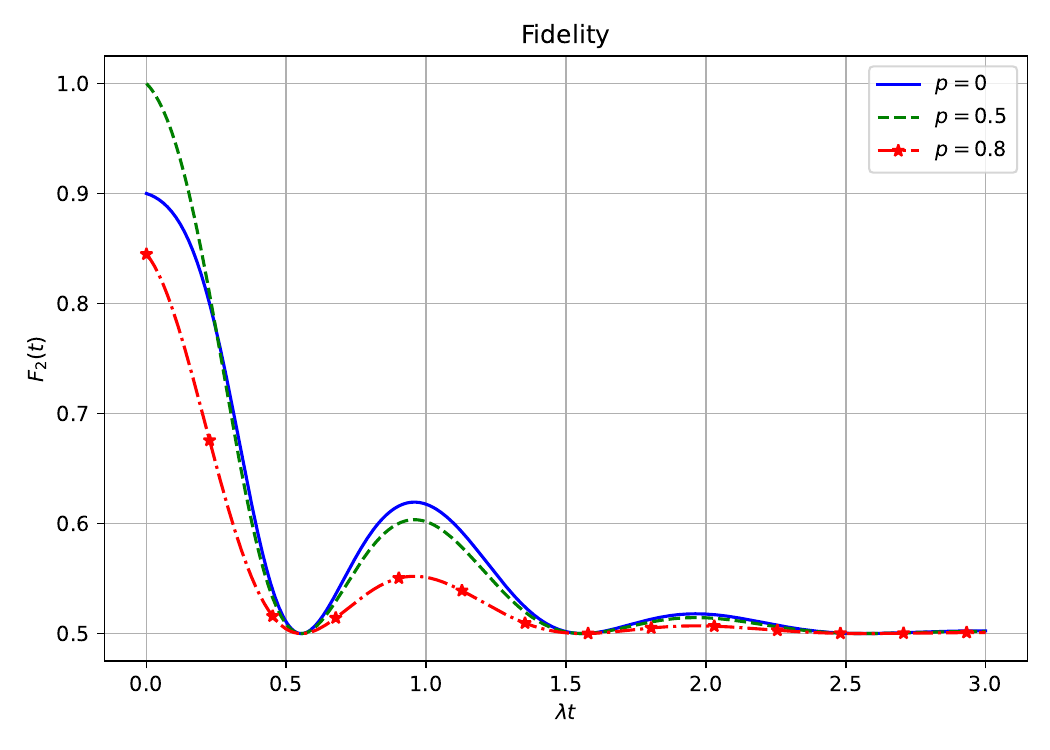} \vfill $\left(a\right)$
 					\end{minipage} \hfill
 					\begin{minipage}[b]{.33\linewidth}
 						\centering
 						\includegraphics[scale=0.33]{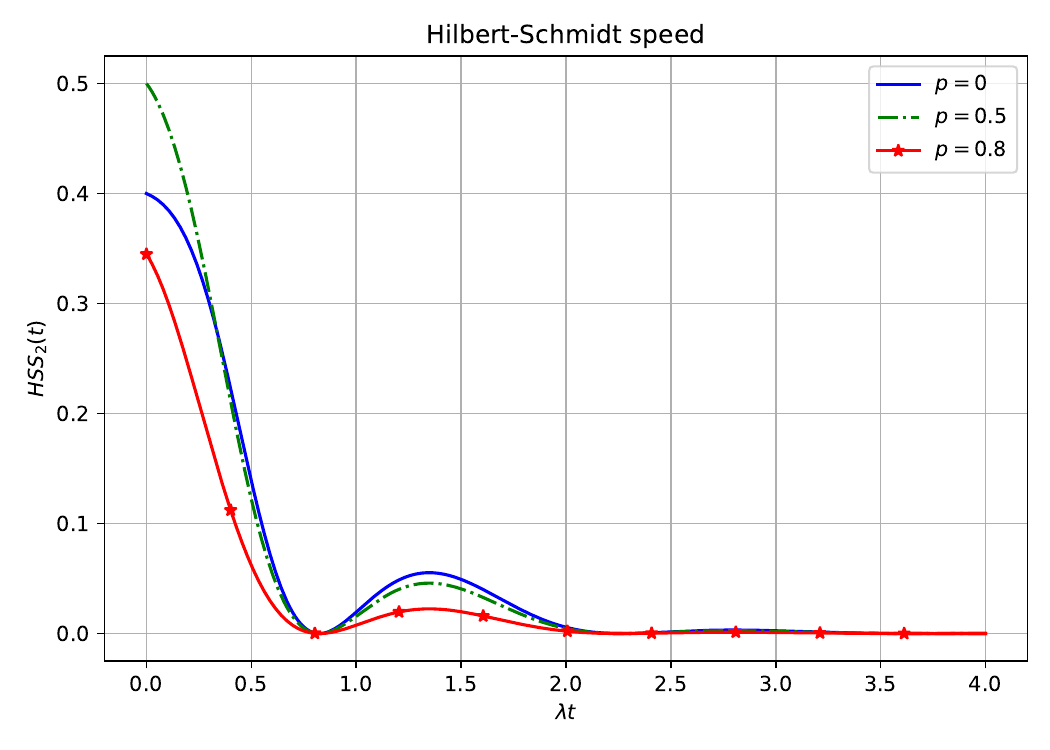} \vfill  $\left(b\right)$
 					\end{minipage} \hfill
 					\begin{minipage}[b]{.33\linewidth}
 						\centering
 						\includegraphics[scale=0.33]{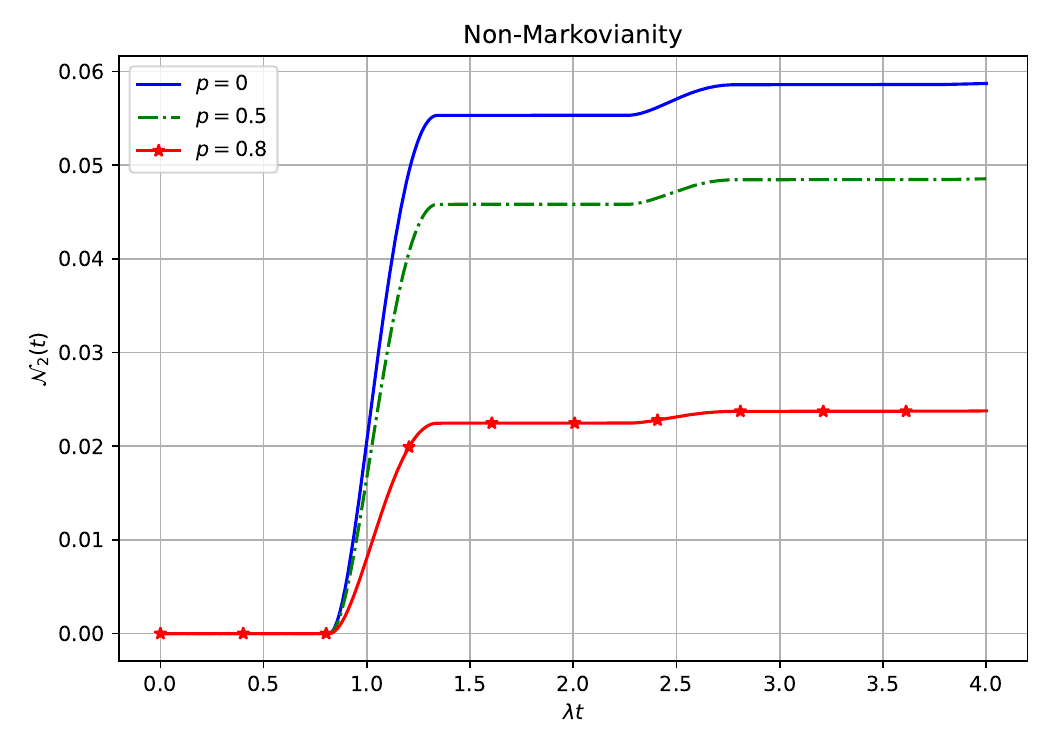} \vfill  $\left(c\right)$
 			\end{minipage}}}
             \caption{Teleportation fidelity $\mathcal{F}_2(t)$,$HSS_2(t)$  and non-Markovianity $\mathcal{N}_2(t)$ as functions of $\lambda t$ with $\gamma_0 = 20\lambda$, for different values of $p$ with fixed $q = 0.5$, and initial state parameters $\theta = \frac{\pi}{2}$ and $\varphi = \frac{\pi}{4}$.}
   \label{fig2}
 		\end{figure}
 	\end{widetext}

Figure \ref{fig2} illustrates the influence of the weak measurement strength $p$ on the system's dynamics, focusing on the fidelity panel (a), the Hilbert–Schmidt speed $HSS_2(t)$, panel (b), and the accumulated non-Markovianity $\mathcal{N}_2(t)$ panel (c), all plotted as functions of the rescaled time $\lambda t$. As the weak measurement strength $p$ increases corresponding to a stronger application of the WM and a smaller post selection probability $\bar{p}$ a significant enhancement in fidelity is observed panel (a). This indicates that the WM+QMR protocol effectively protects the system’s quantum state against decoherence. Interestingly, this enhancement is accompanied by a rapid suppression of dynamical activity, as evidenced by the decreasing and less oscillatory behavior of the $HSS_2(t)$. Concurrently, the non-Markovianity measure $\mathcal{N}_2(t)$ panel (c) diminishes with increasing $p$, suggesting that the backflow of information from the environment is significantly reduced. These observations confirm that the fidelity enhancement is not driven by memory effects but rather by the decoherence suppressing influence of the WM+QMR protocol. Overall, while non-Markovianity can temporarily support coherence preservation, active control via weak measurement and its reversal provides a more effective and tunable mechanism for maintaining long term fidelity in noisy quantum channels.

\section{Enhancing Teleportation Fidelity via Environment-Assisted Measurement and Quantum Measurement Reversal}

In this section, we propose a protocol aimed at improving the fidelity of quantum teleportation under non-Markovian decoherence by employing EAM and QMR. The teleportation fidelity, which quantifies the overlap between the input and the output states, serves as a critical benchmark for assessing the effectiveness of quantum information transfer. Unlike previous sections, where the emphasis was placed on parameter estimation, our focus here is on preserving the overall integrity of the quantum state during teleportation.

We consider a standard teleportation setup in which a maximally entangled Bell state, 
\begin{equation}
|\psi^+\rangle = \frac{1}{\sqrt{2}} (|00\rangle + |11\rangle),
\end{equation}
is initially prepared by a third party (Charlie) to serve as the entanglement channel between the sender (Alice) and the receiver (Bob). Qubits 2 and 3 of this entangled pair are transmitted through local non-Markovian environments to Alice and Bob, respectively. During this distribution, no weak measurement is performed in advance, allowing the entangled state to evolve under environmental noise.

To mitigate the adverse effects of decoherence, we introduce environment assisted measurements. Specifically, detectors are coupled to the environments interacting with qubits 2 and 3. When no excitation is detected a condition associated with the $E_{00}$ Kraus operator the corresponding trajectory is post selected for the teleportation protocol. This selective process leads to a conditional evolution of the entangled state, resulting in the two qubit density matrix
\begin{equation}
\rho_3(t) = \frac{1}{M} \left( \eta_{11}|00\rangle\langle00| + \eta_{14}|00\rangle\langle11| + \eta_{41}|11\rangle\langle00| + \eta_{44}|11\rangle\langle11| \right),
\end{equation}
where $\eta_{11} = \frac{1}{2}$, $\eta_{14} = \eta_{41}^* = \frac{1}{2} \mu(t)$, $\eta_{44} = \frac{1}{2} \mu^2(t)$, and $M = \eta_{11} + \eta_{44}$ ensures proper normalization.

Following the EAM step, QMR is applied to both qubits to probabilistically reverse part of the decoherence effects. The strength of the reversal operation is denoted by \( q' \), and the resulting density matrix after this recovery process becomes
\begin{align}
\rho_4(t) &= \frac{1}{N} \Big(
    \bar{q}'^2 \eta_{11} |00\rangle\langle00| 
    + \bar{q}' \eta_{14} |00\rangle\langle11| \notag \\
    &\quad + \bar{q}' \eta_{41} |11\rangle\langle00| 
    + \eta_{44} |11\rangle\langle11|
\Big),
\end{align}

where \( \bar{q}' = 1 - q' \), and the normalization constant is \( N = \bar{q}'^2 \eta_{11} + \eta_{44} \).

Upon completing the standard teleportation protocol with this recovered channel, Bob obtains the output state:
\begin{equation}
\rho''_{\text{out}}(t) = \frac{1}{N}
\begin{pmatrix}
\rho_{11}(t) & \rho_{12}(t) \\
\rho_{21}(t) & \rho_{22}(t)
\end{pmatrix},
\end{equation}
with the matrix elements given by
\begin{align}
\rho_{11}(t) &= \left( \bar{q}'^2 \eta_{11} + \eta_{44} \right) \cos^2\left(\frac{\theta}{2}\right), \\
\rho_{12}(t) &= \bar{q}' \eta_{14} \sin\theta e^{-i\varphi},
\end{align}
\begin{align}
\rho_{21}(t) &= \bar{q}' \eta_{14} \sin\theta e^{i\varphi}, \\
\rho_{22}(t) &= \left( \bar{q}'^2 \eta_{11} + \eta_{44} \right) \sin^2\left(\frac{\theta}{2}\right).
\end{align}

The teleportation fidelity is then evaluated as
\begin{align}
\mathcal{F}_3(t) &= \langle \psi_{\text{in}} |  \rho_{\text{out}}''(t) | \psi_{\text{in}} \rangle \\ &= \rho_{11}(t) \cos^2\left(\frac{\theta}{2}\right) + \rho_{22}(t) \sin^2\left(\frac{\theta}{2}\right) \\
&+ \sin\theta \, \mathrm{Re}\left[\rho_{12}(t) e^{i\varphi}\right]
\end{align}
which reflects the effectiveness of the combined EAM and QMR strategies in preserving the fidelity of quantum teleportation under non-Markovian noise.

\begin{widetext}

 		\begin{figure}[hbtp]
 			{{\begin{minipage}[b]{.33\linewidth}
 						\centering
 						\includegraphics[scale=0.33]{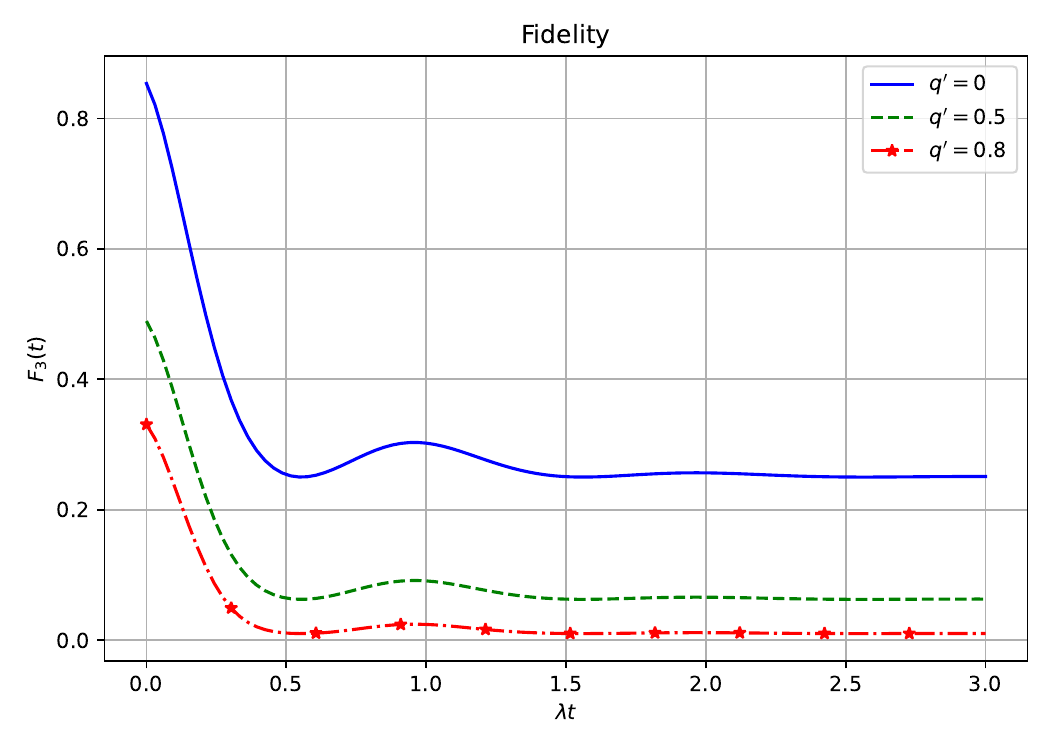} \vfill $\left(a\right)$
 					\end{minipage} \hfill
 					\begin{minipage}[b]{.33\linewidth}
 						\centering
 						\includegraphics[scale=0.33]{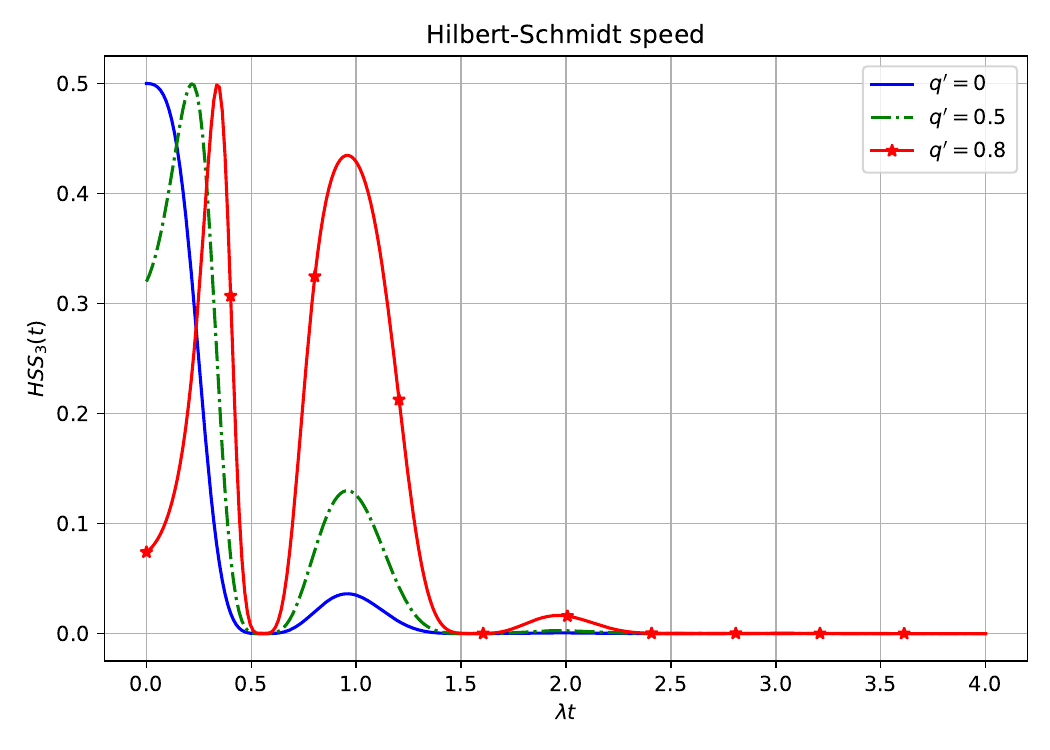} \vfill  $\left(b\right)$
 					\end{minipage} \hfill
 					\begin{minipage}[b]{.33\linewidth}
 						\centering
 						\includegraphics[scale=0.33]{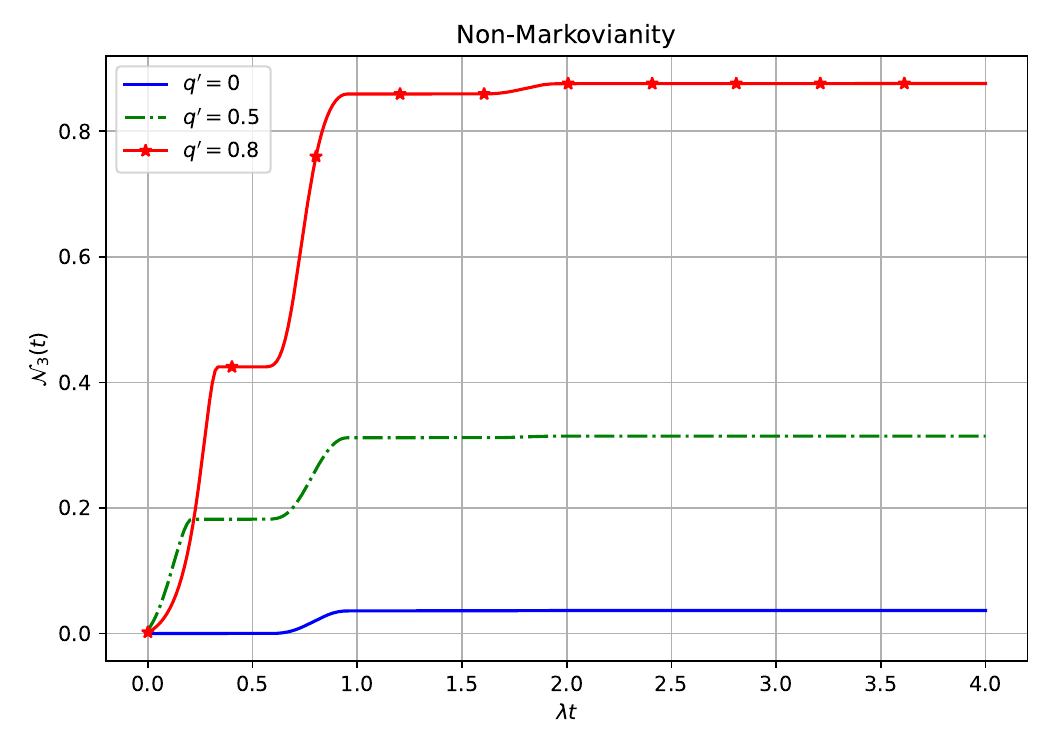} \vfill  $\left(c\right)$
 			\end{minipage}}}
              \caption{ Teleportation fidelity $\mathcal{F}_3(t)$,$HSS_3(t)$  and non-Markovianity $\mathcal{N}_3(t)$ as functions of $\lambda t$ with different fixed values of the QMR strength \( q' \)  . The input state is prepared with \( \theta = \frac{\pi}{2} \) and \( \phi = \frac{\pi}{4} \).  }
    \label{fig3}
 		\end{figure}
 	\end{widetext}
 
 In Figure \ref{fig3}.a, we present the fidelity $\mathcal{F}_3(t)$ under a control strategy that combines EAM with QMR. Different curves correspond to varying values of the QMR strength $q'$, with the initial state fixed at $\theta = \pi/2$ and $\varphi = \pi/4$. This approach probabilistically projects the system into less decohered subspaces by selectively accepting trajectories where no environmental excitation is detected. The subsequent QMR operation further protects the state, resulting in significantly enhanced fidelity compared to uncontrolled evolution.\par

The observed decay in fidelity with increasing $q'$ illustrates a key trade off in the use of QMR. Contrary to intuition, higher values of $q'$ do not enhance fidelity in this scheme; instead, they progressively degrade it. This behavior can be understood by recognizing that QMR becomes less effective as its strength increases due to the probabilistic nature of the post-selection. As $q' \to 1$, the reversal operation tends to distort the post-selected state rather than restoring coherence, leading to reduced fidelity. The robustness of fidelity at lower $q'$ values indicates that moderate reversal strengths are more effective in preserving coherence when combined with EAM. This highlights the importance of fine tuning the reversal strength, as excessive correction can be counterproductive under non-Markovian noise.\par

In panel (b), the Hilbert-Schmidt speed  captures the instantaneous evolution speed. Peaks in the $HSS_3(t)$ coincide with fidelity revivals, indicating that non-Markovian memory effects not only restore coherence but also accelerate system evolution. This supports the interpretation that temporal correlations, inherent to the non-Markovian environment, can be harnessed to achieve faster and more efficient dynamics. The impact of QMR is evident in how the evolution remains dynamically active, even under protective filtering, demonstrating controlled yet responsive behavior.\par

In panel (c), the non-Markovianity measure  quantifies how much information flows back into the system from the environment. The positive, time-localized peaks confirm that memory effects are present and significant. The temporal alignment with fidelity revivals and speed peaks strengthens the interpretation that EAM+QMR benefits directly from environmental memory.\par

Overall, these three panels coherently demonstrate the effectiveness of the EAM+QMR control protocol. By filtering out decohered trajectories and utilizing QMR to reinforce coherent ones, the scheme exploits non-Markovian backflow and entanglement with the environment to sustain and accelerate quantum dynamics. Compared to WM-based methods, it offers better robustness, less parameter sensitivity, and enhanced coherence protection, making it a promising candidate for quantum control in noisy intermediate-scale quantum devices and other realistic platforms where full optimization is challenging.

\section{Conclusion}

In this work, we have investigated how non-Markovian dynamics can be leveraged to enhance quantum teleportation fidelity. Specifically, we analyzed two control strategies based on weak measurement and quantum measurement reversal (WM+QMR), and environment-assisted measurement combined with QMR (EAM+QMR), under the influence of non-Markovian noise.

By using teleportation fidelity as the central figure of merit and HSS as a witness of non-Markovianity, we demonstrated a clear correlation between fidelity revivals and peaks in HSS. This alignment indicates that memory effects characterized by temporary information backflow can indeed restore coherence and improve teleportation fidelity when appropriately harnessed.

Importantly, our results reveal that in the EAM+QMR scheme, increasing the QMR strength $q'$ beyond moderate values can actually degrade fidelity, as the system has already been favorably filtered by EAM. This underscores that non-Markovianity alone, selectively accessed via EAM, can sustain coherence without requiring aggressive correction.

Overall, our findings highlight the operational advantage of non-Markovian environments in quantum information protocols. The EAM+QMR scheme, in particular, offers a robust and less parameter-sensitive approach to preserving fidelity, making it well-suited for implementation in realistic noisy quantum systems.

\section*{ACKNOWLEDGMENTS}
S.G. acknowledges the financial support of the National Center for Scientific and Technical Research (CNRST) through the “PhD-Associate Scholarship-PASS” program. The authors acknowledge the LPHE-MS, FSR for the technical support.

\section*{Declarations} 

{\bf Data Availability Statement:} No data were used for the research described in the article.\par 

{\bf Conflict of interest:} The authors declare that they have no known competing financial interests or personal relationships that could have appeared to influence the work reported in this paper.

\end{document}